# Techniques for the generation of 3D Finite Element Meshes of human organs

**LOBOS, C.[†], PAYAN, Y.[†], and HITSCHFELD, N.[‡]**

[†] *TIMC-IMAG Laboratory, UMR CNRS 5225, Joseph Fourier University, 38706 La Tronche CEDEX, France.*

[‡] *Universidad de Chile, FCFM, Departamento de Ciencias de la Computación, Blanco Encalada 2120, 837-0459 Santiago, Chile.*

## INTRODUCTION

**Continuum mechanics** (CM) is a branch of mechanics that deals with the analysis of the kinematic and mechanical behavior of materials modeled as a continuum, e.g., solids and fluids (i.e., liquids and gases). In a nutshell, CM assumes that matter is continuous (ignoring the fact that matter is actually made of atoms). This assumption allows the approximation of physical quantities over the materials, such as energy and momentum, at the infinitesimal limit. Differential equations can thus be employed in solving problems in CM.

Let $\Omega$ be a volumetric domain defined in 3D and $P$ a point inside $\Omega$. A deformation occurs in $\Omega$ as external (surface or volumetric) forces $f$ are applied to some $P$ in $\Omega$. If $\Omega$ is considered as an elastic body, the deformation is characterized by:

- A displacement vector field $u$ caused by $f$ applied over one or more $P$.
- An internal state of deformation, mathematically defined by a "strain tensor" for each $P$ in $\Omega$.
- The "force reaction" of the body to the external forces; this is mathematically defined by a "stress tensor" for each $P$ in $\Omega$.

The Local Equilibrium of the Medium (LEM) is an expression defined for each $P$ in $\Omega$ that links the displacement, the strain and the stress tensor through Partial Differential Equations (PDE). In order to do so, an analytical relationship is assumed between the strain and the stress tensors, known as the constitutive law of the material.

The difference between displacement and deformation is made in function of the Reference System (RS) used to measure each one of them. The first one is made regarding the overall RS. The second one, uses a relative RS associated to the domain. Figure 1 illustrates the initial state of the system at the top; at the bottom left panel a displacement is produced as the coordinates of point $P$ change to $P'$ regarding the $x$ and $y$ axis, however no deformation is presented in the $r$ and $s$ axis. At the bottom right panel the inverse situation is produced as no displacement of $P$ is presented in the *x* and *y* axis. In the other hand, point *P* has a severe deformation regarding the *r* and *s* axis.

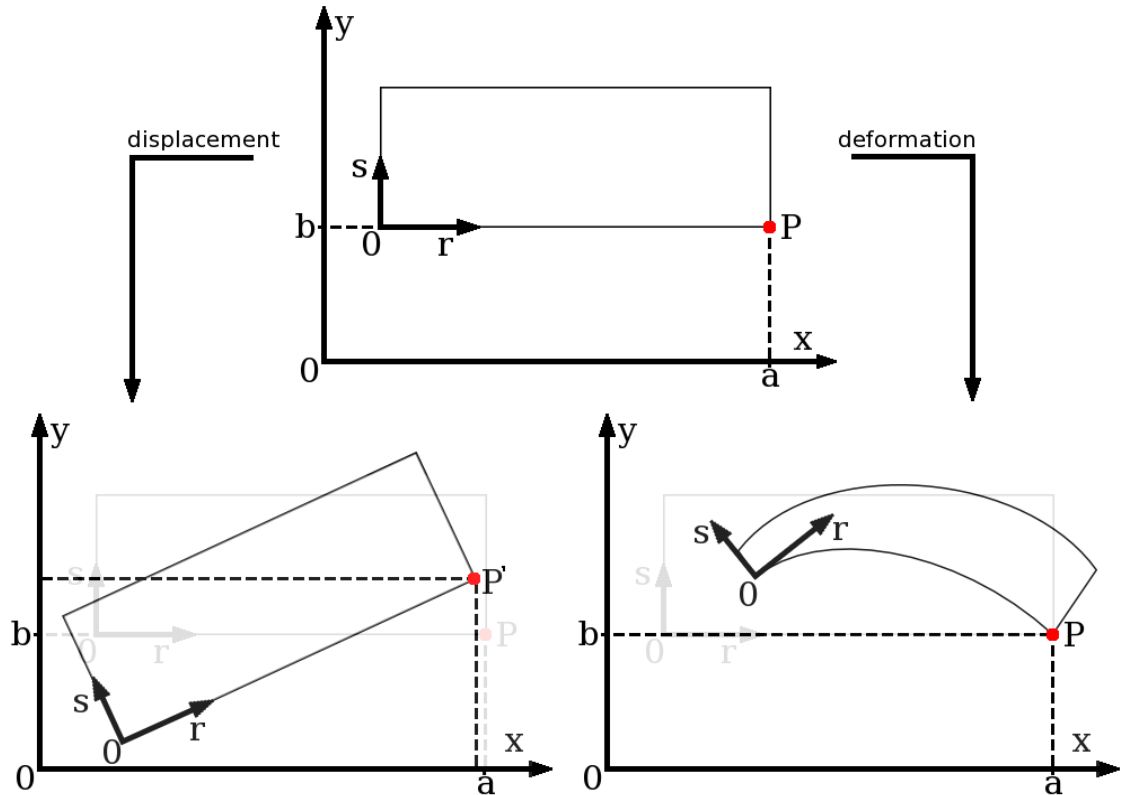

*Figure 1: Displacement vs deformation. Top: Initial system. Bottom left: displacement (a difference regarding the x and y axis). Bottom right: deformation (a difference regarding the r and s axis).*

The strain level measures the domain deformation regarding the relative position of *P* in $\Omega$. Let vector *a* be the initial position of *P* and vector *b* be the position of *P* after the deformation in the relative axis: *r*, *s*. The strain level is then defined as:

$$\frac{|\,\|a\| - \|b\|\,|}{\|a\|}$$

The importance of measuring the strain level is that when it is inferior to 10% the "small strain" hypothesis can be made, which assumes a linear geometrical resolution of the PDEs. Otherwise, a "large strain" framework is necessary, which means a much more complex resolution of the system.

If the constitutive law can be assumed to be linear, another simplification ca be made: an elasticity tensor is computed to link the stress and strain tensors. On the contrary, if a linear law does not account properly for the mechanical behavior of the tissues, a non-linear law can be chosen to account for the hyperelastic behavior of soft tissues (see Bonet and Wood (1997) for more details).

Most of the modeling works proposed in the literature assumed both, the small strain and the linearity of the constitutive law hypotheses. The term "linear elasticity" is then commonly used, i.e. assuming linearity of the geometry and the mechanics. In that case, it can be shown that the constitutive behavior of the material can be characterized by only two parameters: the Young's modulus that depends on the stiffness of the material, and the Poisson's ratio that is related to the compressibility of the material.

Whatever the modeling assumptions are, the PDEs that govern CM cannot be analytically solved over the full domain. They are therefore numerically solved over a piecewise discretization of the domain, usually by means of the **Finite Element Method** (FEM), the Finite Difference Method and the Finite Volume Method. This chapter focuses on the use of FEM since most of the modeling works in the biomechanical field use this method.

The name FEM is due to the principle that the continuous space is subdivided in a finite number of smaller and simpler geometries, namely the elements. This subdivision of the continuous space is called the **mesh**.

## CONSTRUCTING THE SOLUTION OF THE FEM

The **Finite Element Method** (FEM) requires the discretization of the spatial domain Ω with "finite elements" interconnected at points called "nodes". The PDE of the CM can then be solved inside each element which geometry is quite simple and regular. For two dimensional problems, triangles and rectangles are commonly used while in three dimensions tetrahedral and hexahedral elements are very popular. As compared to finite difference methods, the finite element mesh may be entirely unstructured, which makes the modeling of complicated and irregular geometries more convenient.

Altogether, the elements should cover the entire domain as accurately as possible. For a given set of boundary conditions (i.e. forces and constraints applied on nodes) and assuming an analytical constitutive law, the unknowns of the systems, i.e. node's displacements, are solved for each element. Then, thanks to interpolation "shape functions" (usually linear or quadratic), displacements, strains and stresses can be computed at each point inside the elements. Because of element interconnections and continuity of the shape functions from one element to another one, the FEM provides solutions of the system inside the full domain.

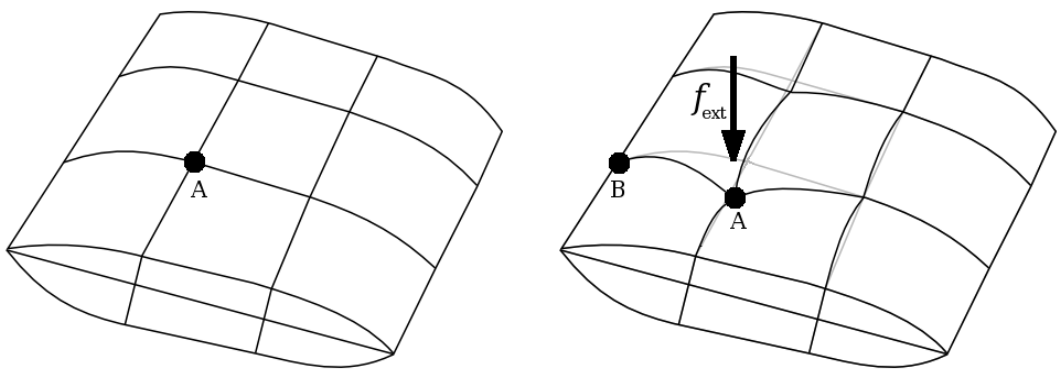

*Figure 2: An example of a deformed domain simulated with the FEM. Left: the initial system. Right: the domain is deformed by the influence of an external force.*

Figure 2 shows an example of FEM over an arbitrary domain. In this example, an external force *f* is applied over node A. Node B is considered as fixed (this is an example of constraint applied over a node). The position of the other nodes in the mesh is obtained in function of *f* and all the modeling parameters in each node (strain, stress and constraints). Finally the position of each point in the domain is obtained by the interpolation function over each element.

## GLOBAL MESHING DEFINITIONS

### What is a mesh?

Several definitions can be found to this question. In this chapter a **mesh** shall be seen as a partition of an arbitrary domain into simpler geometrical objects (or elements). Those elements are compound of nodes, edges, faces and the relations between them.

Probably the most common definition is "a mesh is a tessellation of the space", which is equivalent to our definition since a tessellation is a collection of non-overlapping elements that fill a domain.

Achieving this subdivision of the space is not an easy task, therefore many ways to produce this tessellation have been proposed. Sections "mesh adaptation" and "mesh generation" will give a reference over the most popular and useful ones to the medical domain.

### What is a valid mesh?

The FEM can be seen as an integration problem (a system of PDE). The computation of this integration is produced over the elements that describe the domain. An element is invalid when the area or volume that it describes is malformed. Two types of malformations can be presented in the mesh:
- Edge inversion: this involves several elements and it is produced when the intersection of neighbor elements has a positive volume.
- Concave element: this is a local problem and it is produced when one or more faces of the element are concave (this problem cannot be produced in triangle faces).

In both cases, the area or volume described by the elements is artificially increased and do not represents the current state of the domain causing an integration over a "phantom" sub-domain. The case of "edge inversion" is presented in 2D in figure 3. In this case, the domain is represented by two triangles *t1* and *t2*. Node A, that belongs only to *t1* is dragged into the region defined by *t2*. In this new scenario the domain corresponds to the shaded area. If the mesh defined by *t1* and *t2* is used to represent this new state of the domain, the integration should be over *t2 - t1*.

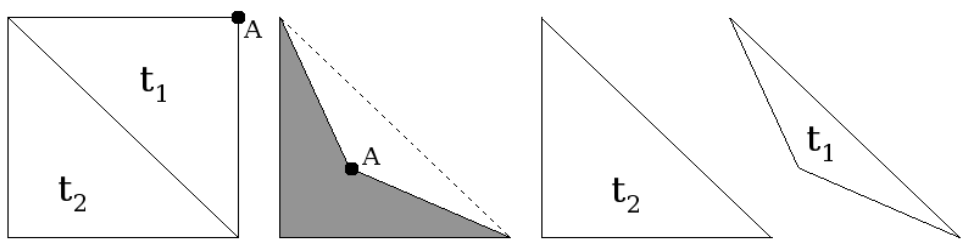

*Figure 3: At left a valid 2D mesh. At right an invalid mesh where shaded triangles have a negative area.*

However, it occurs over *t2 + t1*.

The same problem can be expanded to 3D with hexahedra. Figure 4 shows an example of a concave hexahedron. A valid hexahedron (left) is transformed by the displacement of point *A* (right) in such a way that it turns into an invalid configuration. The rendering of the hexahedron is shown as a prism (wedge) defined by the triangle *B*, *C* and *D* and its projection through the *z* axis. The final numerical integration occurs over the domain described by this prism (while the actual domain is the concave hexahedron).

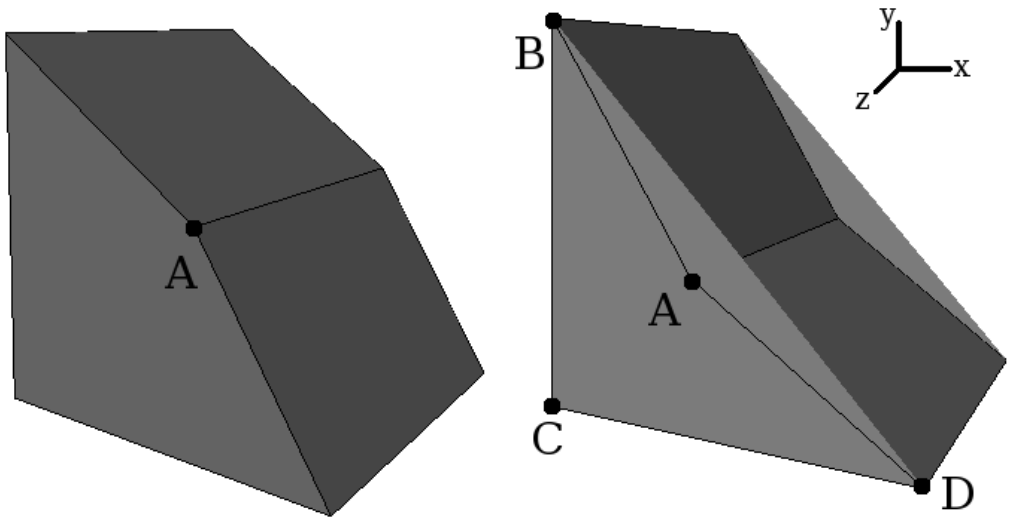

*Figure 4: Left: a valid hexahedron. Right: a concave hexahedron considered as an invalid configuration.*

The Jacobian matrix of an element can be used to determine if an element is valid or not (this will be shown later).

Note that in the literature, a valid mesh can be also named as a regular mesh (see Luboz, Payan and Couteau, 2001).

## What is a good mesh?

Two aspects should be considered to determine if a mesh is good or not. The first has to deal with the representation level of the domain (RLD). This variable is measured as the difference between the areas or volumes of the actual domain and the final mesh.

The second aspect is the quality. The *perfect* element can be described regarding some relations between the angles, edge's length, distance between specific element's points, circumcircle, etc. Unfortunately, no "magic" quality measurement exists for an element, because it depends on the numerical method been used and on the problem being solved.

In the case of tetrahedra, the work of Shewchuk in 2002 is probably the most relevant article concerning mesh quality. Note that in his web page[1] an unpublished article of 66 pages (just about tetrahedrons quality) can be found. The idea of putting a lot of emphasis in quality is because bad quality elements can lead to several computational errors in the simulation.

In order to illustrate some quality measures, let's describe briefly some classical quality criteria, namely the *aspect ratio*, the *warping factor* and the *dihedral angle*.

---

1 http://www.cs.berkeley.edu/~jrs/

To obtain the Aspect Ratio (AR) of an element, the distances between element's faces must be computed. The AR for the element is then defined as the ratio between the minimal and the maximal distance. Therefore, an AR = 1 corresponds to an ideal element and as the AR reaches high values, the element becomes increasingly distorted (see Lobos, Bucki, Hitschfeld and Payan (2007) for more details).

The Warping Factor (WF) is a quality measure over the element's faces. For each face, the distances of the face's nodes to an average plane are computed. If all the nodes are co-planar, then the WF is 0 and the face is said to be "perfect". As the WF increases, the quality of the face (thus the element) decreases. Note that WF is out of context for triangle faces as 3 nodes are always co-planar.

The Dihedral Angle (DA) is the angle formed between two planes. Therefore several DA values are computed for each element (all the angles between connected faces). The DA of the element is the "worst" value among all the calculated DA. Note that for each type of element the optimal DA is different: 60° for the tetrahedron and 90° for the hexahedron. Therefore the quality

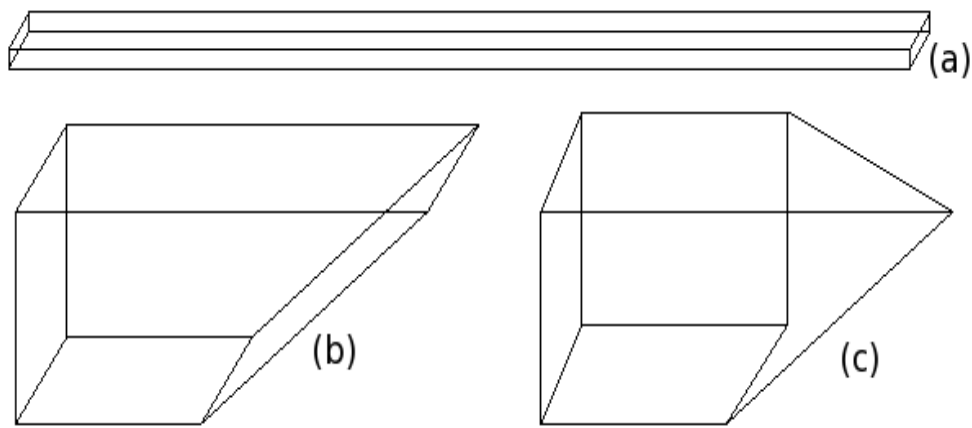

*Figure 5: Bad elements regarding (a) aspect ratio, (b) dihedral angle and (c) warping factor*

of the element decreases when the "worst" DA of it moves further from the optimal value.

Note that improving some of these quality measures doesn't necessarily produce a "good" element. For example, if only WF and DA are considered as quality measures then the element in figure 5 (a) would be considered as "perfect". However in terms of AR, this element has a very poor quality. Following the same principle, figure 5 (b) has a poor DA but a perfect WF and a good AR, and (c) has a poor WF and acceptable DA and AR.

**CHOOSING THE TYPE OF ELEMENT**

In this section the advantages and disadvantages of using the different types of elements in a mesh are presented. Three categories will be analyzed: tetrahedral, hexahedral and mixed-element meshes. As it will be show in the following sections, most of the meshing techniques proposed in the medical field produce tetrahedral meshes. Some works are done with hexahedra and just a few with mixed-elements.

As this chapter is focused in medical simulations, many works in this field consider deformable domains. A very important work over model deformations has been done by Benzley,

Perry, Merkley, Clark and Sjaardema (1995) from the mechanical point of view. This work showed the differences obtained by using tetrahedral or hexahedral meshes in terms of incompressibility and plasticity abilities of each type of element. In this study a simple bar, fixed at one end, with a rectangular cross-section was used to compare the performance of linear and quadratic displacement assumption over tetrahedra and hexahedra meshes. The linear or quadratic property of an element has to deal with the manner to represent their edges and in consequence, the volume the element covers. As figure 6 shows, in a linear element the edges are represented as an interpolation of the two border points. In the case of quadratic elements, the edges are quadratic functions and therefore an additional middle point is necessary to describe the function. If linear tetrahedra have 4 points, its quadratic counterpart has the same 4 point plus 6 others located on each edge of it.

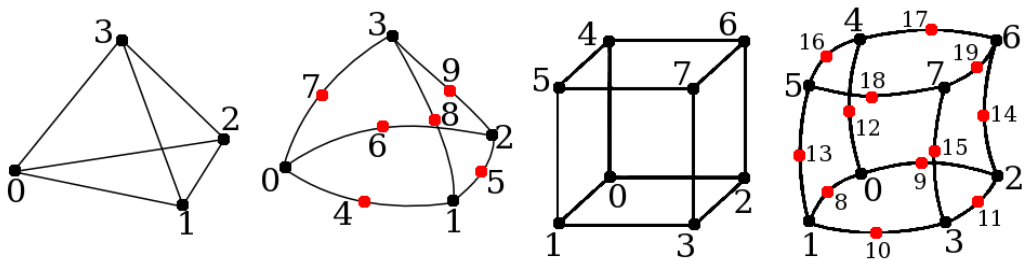

*Figure 6: From left to right: Linear Tetrahedron, Quadratic Tetrahedron, Linear Hexahedron and Quadratic Hexahedron*

The work from Benzley et al (1995) has several important conclusions:
- Linear hexahedrons (LH) can generally deform in a lower strain energy state (i.e. eigenvalues), thus making them more accurate than linear tetrahedrons (LT) in numerous situations.
- The comparison of linear static bending[2] situation indicated that LT models produced errors between 10\% and 70\% in both displacement and stress calculations. Such errors are obviously unacceptable for stress analysis work. On the contrary LH, quadratic hexahedrons (QH), and quadratic tetrahedrons (QT) models all provided acceptable results, even with relatively coarse meshes.
- The linear static torsion problem[3] again showed that the LT element produced errors of an unacceptable magnitude. This problem also demonstrated that, because selective integration is only effective on the bending problem, the LH element, without a significant number of degrees of freedom, produces poor results. Here, as in the previous problem, the QH element is superior.
- Significant information is conspicuous in the nonlinear elasto-plastic calculations. Here, as before, all but LT models are adequate for bending calculations. However, not only LT, but QT models seemed to underperform both LH and QH elements.

As it will be shown, most of the meshes used in medical simulations only consider linear elements. There are just a few works that use quadratic elements, even taking into account that some commercial Finite Element (FE) solvers like ANSYS® are prepared to manage those kinds of elements. In other words, there exist FEM solvers that manage quadratic elements, however and even when this may lead to better simulation results, there are few mesh generators that produce quadratic elements.

---

[2] Evaluate linear elements in a single new state after bending the bar.
[3] Evaluate linear elements in a single new state after some torsion is applied over the bar.

Other type of variables can be taken into consideration to choose the elements in the mesh. For example, regarding linear elements, only triangular faces will always be planar. Once again, commercial FE solvers like ANSYS® have specific and well defined thresholds to accept those kinds of elements. Even though making a non-triangle face "more" planar has a solution, this is an extra problem to consider while using other type of element than the tetrahedron (or a dual element taken from it).

Finally, there are some other works that have studied the use of different types of elements (Lobos et al, 2007; Couteau, Payan & Lavallee, 2000). The most global definitions consider tetrahedra, pyramid, prism (wedge) and hexahedra. Two options are possible to consider this kind of meshes:

- A better approximation of the domain in a given region of it. This is the case for meshes that are governed by one type of element, but that prefer in some very specific configurations, to replace bad quality elements with another type of elements.
- Transitions between different levels of refinement. Even though this can be achieved by using just one type of element, some works prefer to include different types of elements during transitions between coarse and refined regions.

The "locking effect" could be another drawback to the use of tetrahedra for quasi-incompressible materials (refer to Hughes (1987) for more details).

**IMPORTANT REMARKS OVER MESHING**

Validity (or mesh regularity) should be clearly differentiated from the mesh quality. The element's quality optimization[4] involves node displacement. This last procedure might cause neighbor element invalidation. Therefore quality optimization should be always made considering not affecting the validity of the rest of the elements. For instance figure 7 shows how the displacement of point A could be considered as an action to improve the top left triangle's quality.

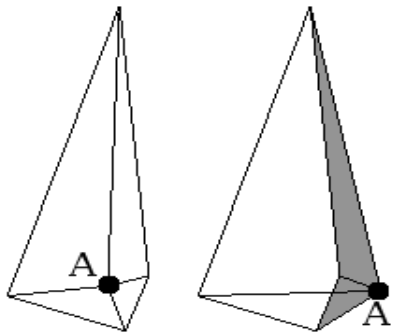

*Figure 7: Right: a valid 2D mesh. Left: an invalid mesh where shaded triangles covers an area that is two times considered in the numerical integration.*

---

[4] In the meshing field, optimization refers to mesh quality improvement without inserting new nodes.

Unfortunately this causes the invalidation of the two right triangles and makes the entire mesh useless (it cannot be used to produce a simulation with the FEM).

As mentioned before, the representation level of the domain (RLD) is the difference in area or volume between the domain and the final mesh. Another remark can be made relating the RLD and the mesh quality. Figure 8 (a) shows an arbitrary domain. In (b) a tessellation of the domain is presented and the RLD is shown by the shaded regions. In (c) some triangles of (b) were split by the addition of new points which makes the RLD much more precise. In figure 8 (d) shaded elements correspond to the triangles of (b) that were split resulting in the tessellation presented in (c). As mentioned before, one of the quality measures is the Dihedral Angle (DA) for 3D and just the angle for 2D. If this quality measure is applied to the final mesh in figure 8 (d) light shaded triangles would be considered as "bad triangles" (because the minimal angle is less than a threshold value) and dark shaded as "good". Following the "angle" quality parameter, the overall mesh quality in (b) would be better than the one presented in (c). Therefore one important remark is that improving the RLD does not necessarily increase the quality of the elements and vice-versa.

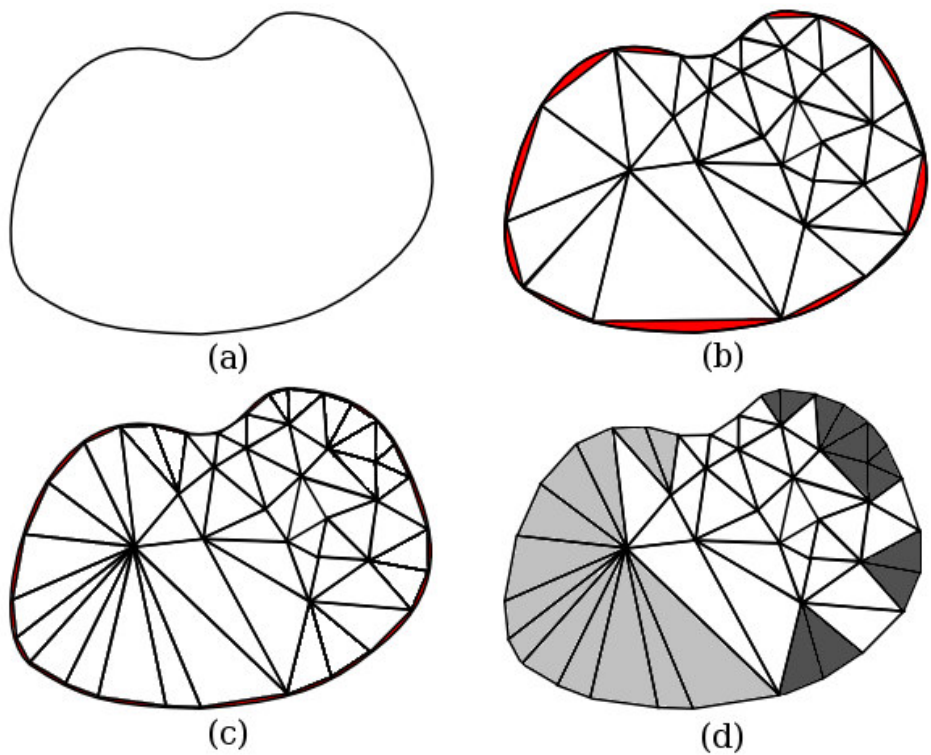

*Figure 8: (a) input domain; (b) a mesh of the input domain where shaded regions represents the RLD; (c) some elements of (b) were split which causes an improvement of the RLD; (d) the mesh from (c) where the split triangles from (b) are shaded. The minimal angle quality measure is applied over shaded triangles and light ones are "bad" while dark ones are "good".*

As shown through the chapter, several factors must be considered before building a mesh. In particular the following questions should be answered:
- should we use one type of element or several? which ones?
- is it desirable to count with different levels of refinement through the mesh?

- how quality will be measured?
- how quality improvement will be achieved when necessary?
- how invalid elements will be repaired?

Once answers are provided for those questions, the next step concerns the way the mesh is going to be built.

## INPUTS FOR THE GENERATION OF HUMAN ORGANS FE MESHES

In medical applications information concerning the domain to mesh comes from one or more of the following inputs:
- Computed Tomography (CT) images.
- Magnetic Resonance Images (MRI).
- Segmented images.
- A surface mesh.
- A cloud of points.
- A pre-generated volumetric FE mesh that describes a generic organ.

While CT is more adapted to bone visualization, MRI is better to differentiate soft tissue from an image. Both techniques can be the starting point for the elaboration of a 3D organ geometry. As a person lies in a scanner, several slices of a body region are captured in images. Figure 9 shows how it is possible to build a surface mesh from this set of piecewise 2D images. Pages, Sermesant and Frey (2005) describe one alternative to do the whole process: from image to surface and volume meshes.

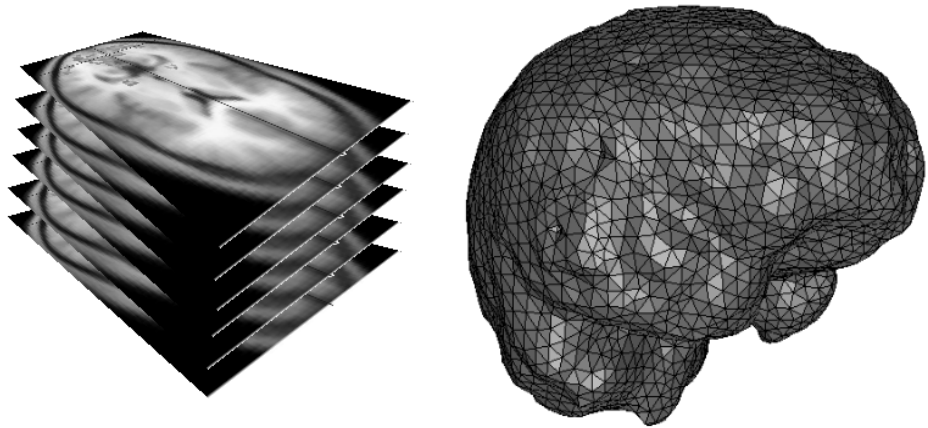

*Figure 9: Left: an example of 2D scan images of the brain. Right: a surface triangle mesh.*

A more elaborated input can be the same set of images but with already segmented organs. This process produces a partition of a digital image into multiple regions with objects and boundaries (points, lines, curves, etc) in the images. Several general-purpose algorithms and techniques have been developed for image segmentation. Since there is no general solution to this problem, these techniques often are combined with domain knowledge, i.e. semi-automatic process leaded by the user to identify the target structures.

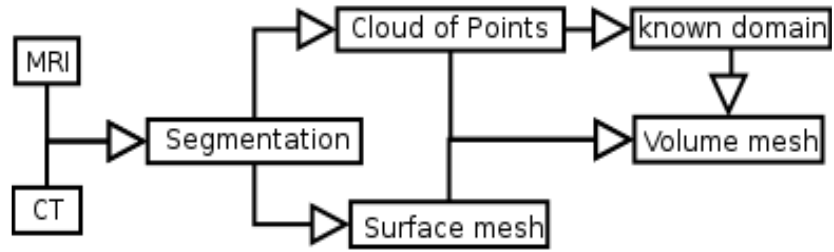

*Figure 10: The different alternatives to go from imaging to volume mesh generation.*

Subsequently, from the segmented images, a cloud of points can be extracted or a surface model can be constructed. In order to produce a surface mesh, algorithms like the Marching Cubes (Lorensen & Cline, 1987) are very popular (see later for more details).

Most of the meshing techniques that will be explained in further sections start from a surface model of the organ. However one family of meshing techniques does not: a cloud of points is enough as they have additional information on the target organ to mesh (see "mesh registration" section for more details).

Figure 10 shows the different paths to produce a volumetric FE mesh from a set of images and further steps. Note that from a cloud of points, it is also possible to build a triangulation and therefore a surface mesh.

## MESHING AND SIMULATION CLASSIFICATIONS

### Global classification of meshing techniques

When the geometry to mesh is known a priori there is a substantial advantage because there is more information that can be used. For example, if the target geometry to mesh is a femur, a pre-calculated mesh (ATLAS) that describes a generic femur can be build. Now, to mesh any other femur, only the external surface information of this new target is necessary and by causing nodes displacement over the ATLAS, it is possible to fit the target information. In other words, it is possible to start from a valid pre-defined FE mesh solution and then deform it in order to fit the target's border conditions.

Consider now the case where anatomically malformed femurs have to be modeled. The previous described technique may not work as it is constrained to "small differences" between the ATLAS and the target. If this constraint is not respected, the elements of the resulting mesh can be strongly degenerated or present bad qualities, which will decrease the accuracy of the simulation. Therefore for those types of problems (unknown domain and big changes from one target to another) it is recommended to use other types of meshing techniques.

In most cases those techniques start from an input surface mesh and create a new target-specific volumetric mesh. The other option is to use the surface information to know if a point is outside or inside the geometry boundaries. In those cases, the resulting mesh will not depend on the properties (density, types of elements) of the input surface mesh. This last family of techniques can also be adapted to use a cloud of points instead of a surface mesh; however the resulting mesh can be not so accurate in those zones where point density is not high enough.

Regarding all this, we can now identify two main streams to generate a volumetric mesh:

- **Mesh adaptation**. Those techniques start with a pre-defined FE mesh ATLAS and then, by moving the ATLAS nodes, achieve the representation of a new target domain.
- **Mesh generation**. Those techniques produce an *ad-hoc* volumetric mesh from the specific input domain.

Both streams will be detailed later in the chapter.

## Simulation types

In general, two types of simulations can be distinguished: the ones focused on obtaining fast / interactive results and the ones focused on precise / accurate results. We could also say that some techniques focus on both fast and precise results, but in real practice, most of the times we have to make a choice.

It is very important to mention the differences between fast generation and fast simulation in the meshing field. The first refers to the time needed to produce a mesh. The second refers to the time a numerical framework, like the Finite Element Method (FEM), needs to compute the simulation results.

Regarding the meshes, two aspects must be analyzed. The **quantity of elements** in the mesh is an important variable. The more elements the mesh has, the more the time the FEM needs to produce the simulation results. In the other hand, with more elements it is possible to achieve a better representation of the domain by increasing the quantity of elements in the zone where higher geometry accuracy is required.

The other important variable to achieve a good result is the **quality of the mesh**, however this will increase only the precision of the results. If fast results are required, three options seem possible:

- Fewer elements.
- The use of a cluster of powerful computers with parallel algorithms.
- Some modifications of the FEM with proposals for other modeling and numerical methods (Schwartz, 2001; Picinbono, Delingette & Ayache, 2003; Cotin, Delingette & Ayache, 1996; Nesme, 2008).

## BASIC TOOLS AND PROPERTIES OF MESH GENERATION

## Delaunay property

In 2D, a **Delaunay** triangulation (Delaunay, 1934) for a set $P$ of points in the plane is a triangulation $DT(P)$ such that no point in $P$ is inside the circumcircle of any triangle in $DT(P)$ (except for the three points of $P$ that conform the triangle). In 3D, the concept remains the same, replacing the circumcircle by the circumsphere. In particular, 2D Delaunay triangulation maximizes the minimum angle of all the triangles in the tessellation.

This is a mathematical property, therefore there is no specific meshing technique associated to it. We can only say that a triangle or a tetrahedral mesh satisfies or not the Delaunay property. In the rest of this chapter, a Delaunay mesh is said to be a mesh that satisfies the above property.

Several meshing algorithms start from the Delaunay mesh because it gives an initial quality value to the mesh. This quality value can be improved by moving, adding or removing some points, achieving a better representation of the input domain.

If a triangle does not satisfy the Delaunay property, the flipping edges technique (see Shewchuk, 1999) can always repair this problem in 2D. This is not true for 3D. Figure 11 illustrates this technique. At the left is a triangulation for a set of points that do not satisfy the property. At the right, the same points are plotted but now with a different triangulation that satisfies the property by changing the common edge of the triangles.

As mentioned before, a Delaunay mesh is a reference in the field of tetrahedral meshing techniques. A lot of works stand over the Delaunay property. However it is very important to mention that the Delaunay property is a "good" starting point but not sufficient to produce quality meshes. Sometimes, a Delaunay mesh can lead to numerical instabilities and that's why it must be refined in order to accomplish other quality constraints. Probably one of the most used refinement technique to produce high quality meshes was proposed by Ruppert in 1995. An implementation, freely available on the web, was implemented by Shewchuk (2002a) in his application Triangle with some improvements over the quality of the output mesh. Just to mention one of this improvements, Miller, Pave and Walkington (2003) continued with the work by improving the triangle's quality in terms of minimal and maximal angle.

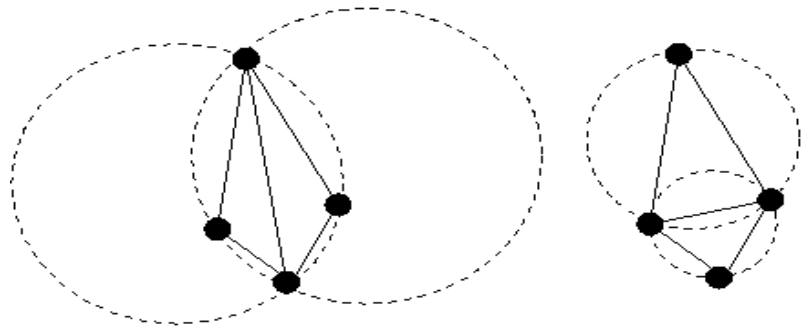

*Figure 11: The Delaunay property. Left: the property isn't achieved. Right: using the flipping edge technique, the property is satisfied.*

To be clear about it, the workflow in most cases is to produce an initial triangle mesh from a cloud of points. The second step is to achieve the Delaunay property for every triangle in the mesh. Finally, the algorithms proposed by Ruppert (1995), Shewchuk (2002a), Miller *et al*. (2003) and many others, improve the quality of the Delaunay mesh by moving, adding or removing some points.

## Grid, Voxel and Octree meshes

A **grid** (or structured) mesh is a regular tessellation of a given space. It is usually formed using a unique type of elements (squares, rectangles, triangles, etc) that can be:
- Equal in angle: all the elements have the same inner angles but not the same area or volume.
- Equal in area or volume: elements can differ in their inner angles.

A voxel is a small volume element, representing a value on a regular grid in the three dimensional space. This is analogous to a pixel, which represents 2D image data. Voxels are frequently used in the visualization and analysis of medical and scientific data. A voxel mesh is a particular grid where all the elements are hexahedra of the same size.

Voxel meshes are particular useful when the simulation is performed over a 3D imaging exam and virtually all the voxels can be activated. For instance in the context of the brain activity study (functional MRI for example), works like Ashburner and Friston in 2000 and Abell *et al.* in 1999 (just to mention some) allow to identify the regions of the brain that are activated and to compare between different groups of subjects.

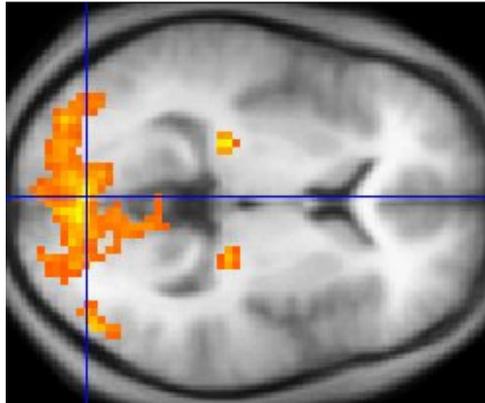

*Figure 12: An example of a voxel mesh over head MRI. The voxel in this case, measures the level of activity over different zones of the brain.*

In those studies, voxels in the mesh allow to identify not only if there is activity in certain region but also, with a pre-defined scale, the level of activity as figure 12 shows[5]. Here the level of the mesh refinement is crucial to detect with precision the regions that are activated. In the other hand, there is a direct relation between the number of voxels and the computing time to process the information. Therefore, finding the adequate level of element's density (the size of each voxel) is crucial.

An **octree** is a volumetric mesh and it can be seen as a derivate product of the quadtree in 2D. A quadtree is a tree data structure in which each internal node has up to four children. Quadtrees are most often used to partition a two dimensional space by recursively subdividing it into four quadrants or regions. The regions may be square or rectangular, or may have arbitrary shapes. This data structure was named a quadtree by Finkel and Bentley in 1974.

Figure 13 shows an example of the tree structure for an octree. Every cube (or octant in this case) is divided, when necessary, in 8 new octants. In the figure, the initial octant numbered 1 was divided in the octants from 2 to 9, then the octant numbered 3, was divided in octants from 10 to 17. Octants 8, 9, 16 and 17 are not visible.

To stop the refinement (the subdivision criterion) on a quadtree or octree, there are two options:
- Geometrical constraint. Like a minimal distance between octree "surface" points and the actual surface of the domain.

---

[5] Image taken from wikipedia http://commons.wikimedia.org/wiki/Image:FMRI.jpg with a Public-domain license. Page last visited: July 21, 2008

- A level constraint. Like to reach a maximum number of subdivisions, points or elements in the mesh.

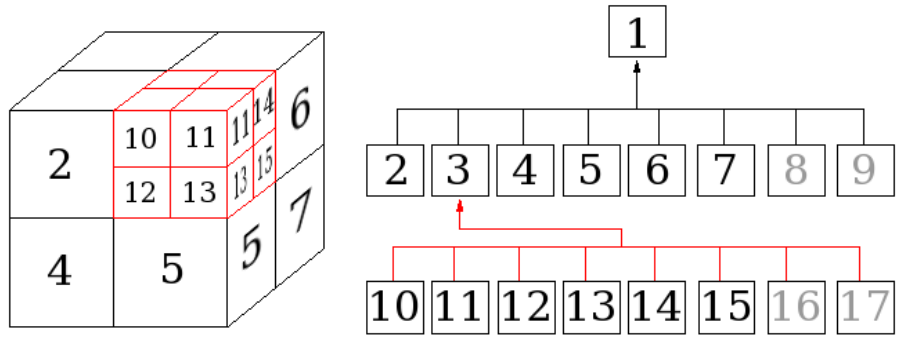

*Figure 13: Left: an example of mesh generated with the octree algorithm. Right: the corresponding tree data structure of it.*

Figure 14 shows an example of octree mesh from a prostate surface mesh. At the left the input surface mesh of a prostate is shown in solid and transparent elements correspond to the octree. Half of the octree mesh has been deleted to show the interior of it. At the right, the same meshes are plotted but now with the input prostate mesh in transparent. Colors in the octree mesh (at the right) show the different sizes of inner elements that can be found. This difference is explained as "big" inner elements don't intersect the input prostate surface mesh and therefore do not continue the splitting process.

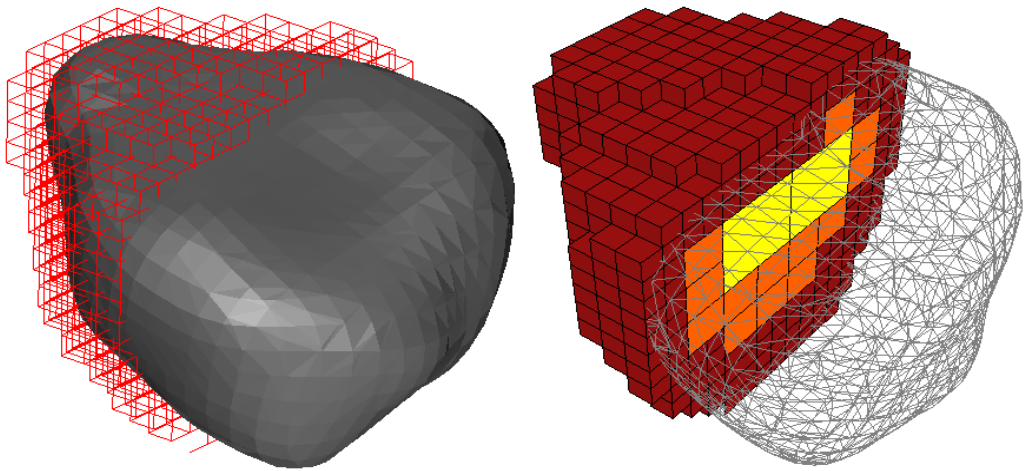

*Figure 14: An example of octree mesh from a prostate. At the left the octree mesh is shown in transparent while the input prostate surface mesh is shown in transparent in the right panel.*

The previous description includes the "pure" octree technique, but in order to make this mesh suitable for FEM it is necessary to manage the transitions between coarse and more refined regions. Here three options are possible:
- Manage transitions using only hexahedra (Zhang and Bajaj, 2006).
- Manage transitions using different types of elements (tetrahedra, pyramid and prism or wedge), (Hitschfeld, Navarro and Farias, 2000; Hitschfeld 2005).

- Do not manage transitions and modify the classical FEM (Nesme, 2008).

The two first alternatives use templates (patterns) to identify different types of configuration and add elements avoiding the refinement of the entire mesh to the same level. These two alternatives also need a technique in order to achieve better surface representation, like a projection of all the nodes that reside outside the surface into it. This can lead to distorted elements and has to be handled by a reparation algorithm. The other option to achieve surface representation is the largely used marching cube algorithm that will be presented in the next subsection.

The third alternative doesn't need any further changes from the current state of the mesh. The effort must be made over the FEM in order to accept elements with points inserted in their edges. The PhD thesis of Nesme (2008) describes how this can be done.

## The Marching Cubes Technique

**Marching Cubes** is a technique published by Lorensen and Cline in 1987 for extracting a polygonal mesh of an isosurface from a three-dimensional scalar field (sometimes called voxels).

The algorithm proceeds through the scalar field, taking eight neighbor locations at a time (thus forming an imaginary cube), then determining the polygons needed to represent the part of the isosurface that passes through this cube. Since there are 8 points in each cube and every point can be in two states, inside or outside the isosurface, there are $2^8 = 256$ different combinations. However, most of the combinations are topologically equivalent and by applying permutations, rotations and switching states of inside/outside, these combinations can be reduced to 15 patterns as shown in figure 15.

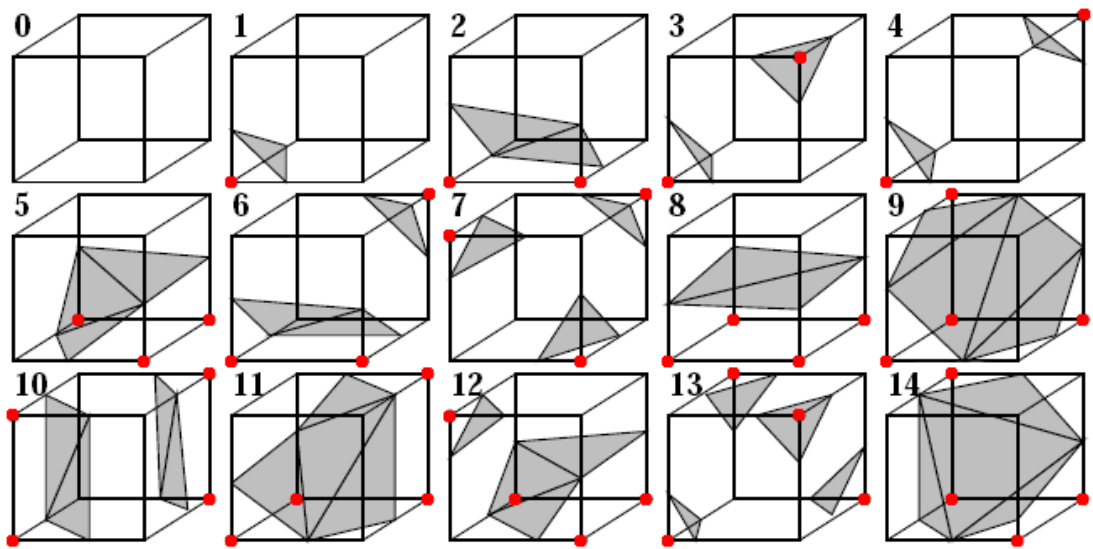

*Figure 15: The 15 marching cubes patterns. Red points can be inside or outside the target isosurface to represent. The resulting faces will represent the isosurface.*

For the switching states, if there are 7 points inside or just one of the entire cube, the same pattern (number 1) will apply to find the intersection face. Even though this procedure is robust, it doesn't work all the time as it doesn't consider neighboring information. The Marching Cubes technique can produce holes as described and solved by Chernyaev in 1995. These holes are the result of combining two different configurations for neighbors as figure 16 shows.

The other option to solve the above problem is to produce a tessellation of the cell into tetrahedra and build the isosurface applying the marching tetrahedra algorithm (Payne & Toga, 1990; Ferrant *et al.*, 1999). This method is similar to marching cubes but based on a tetrahedron instead of a cube. The advantage is that in this case, no hole can be produced. Unfortunately the topology produced by this new process isn't necessarily the same as the one obtained without tetrahedra tessellation (Velasco, Torres, Leon & Soler, 2007).

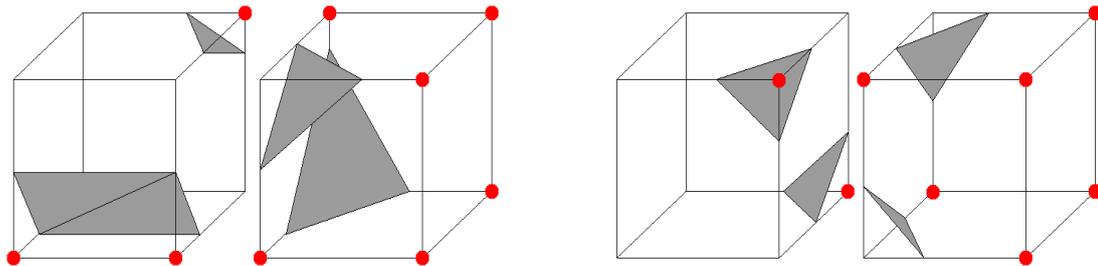

*Figure 16: Two case of erroneous surface representation, by applying the patterns between two neighbors: a hole is produced in the element's shared face. Red points are outside points. Left: pattern 6 and the complement of pattern 3. Right: pattern 3 and its complement.*

The Marching Cubes and their improvements are not only used with the voxel meshes, but can be combined with the octree and grid meshes. This is a fast and well studied technique to improve the surface representation of an input domain.

## Advancing Front Technique

The **Advancing front** is a 2D and 3D meshing technique that stands from a surface (or in 2D a boundary) representation of the domain. Each face in the surface mesh is considered as a front. The idea is to expand all the fronts into the inner part of the volume until the entire domain is meshed.

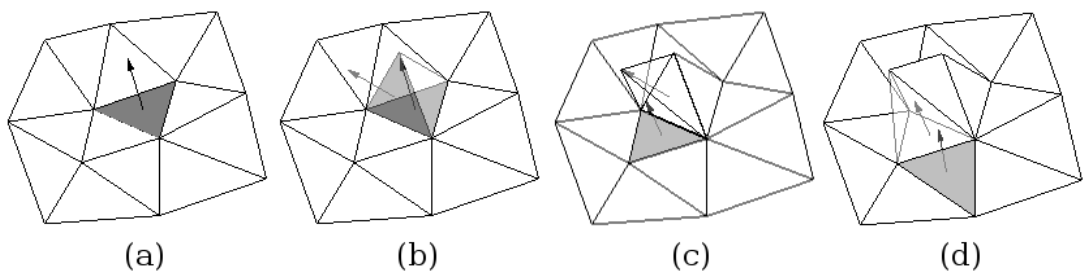

*Figure 17: The advancing front technique: (a) A portion of a surface mesh with one front to expand, (b) the tetrahedron is created and the new faces can be treated as fronts, (c) another expansion using a recently inserted front and (d) another expansion using already inserted points.*

Once a front is expanded, it is no longer considered as a front. Finally all shared faces will no more be fronts as they cannot be expanded. The selection of points to create the new faces (thus elements) encourages the use of existing points. Figure 17 illustrates how the advancing front works.

Note that this technique can be adapted to produce any type of elements (tetrahedra, hexahedra and mixed-elements in general: Lobos & Hitschfeld, 2006). Also note that there is no constraint over the technique to insert a new point. There are several advancing front - based techniques and the difference between them is the manner to chose where and when inserting a new point. This last step tries to maximize the quality of the elements in the mesh. As there are several options to measure the quality, this technique has several flavors.

In the next two sections more complex meshing strategies are presented. Several among them, use the techniques described above as a starting point or as part of the meshing process.

## MESH ADAPTATION

### The mesh-matching algorithm

Couteau *et al.* in 2000 proposed the so called Mesh-Matching (M-M) algorithm. The statement of this work is that building a "suitable" mesh for their purposes is a complex task: count with just a few elements and achieve a "good" domain representation is always a difficult compromise. Therefore in many cases, meshes are built by hand. As this task is time consuming and cannot be done in practice for every patient, the idea is to use a pre-defined FE mesh of the target domain in order to achieve its representation.

The pre-defined mesh is called the ATLAS. This mesh is, in most cases, built by hand in order to preserve element quality, orientation and density in the desired regions, in terms of what is important for the simulation.

To build a new mesh, target information must be represented with a cloud of points at the surface of the new domain to mesh. Then, a registration process *matches* surface nodes of the ATLAS with the target surface points. The goal of this process is to find a 3D transform $T$ that is the combination of a rigid-body transform $RT$, a global warping $W$ and local displacement function $S$ as follows:

$$T_P = RT \circ W \circ S$$

Where $p$ is a vector gathering the parameters of $RT$, $W$ and $S$. Let $M = M_i$, $i = 1,…, N_1$ and $P = P_i$, $i = 1,…, N_1$ be respectively the ATLAS surface nodes and the target surface points (for example patient data obtained by the segmentation algorithm). The elastic registration algorithm minimizes a least-squares criterion $E(p)$ given by:

$$E(p) = \sum_{i=1}^{N_1} \frac{1}{\sigma_i^2}[dist(P, T_p(M_i))]^2 + R_p$$

where $R$ defines a regularization term that is applied to $S$ to obtain a smooth displacement function, $\sigma_i^2$ is the noise variance of the measurement $i$ and *dist* is the distance between $P$ and a point $M'_i$ (transformed by $T$).

As $T$ is a 3D transformation function, it can be applied to any point in the initial model. When $T$ is applied to the internal nodes of the ATLAS, a final 3D target mesh is achieved. The matching of a domain can be seen in figure 18 where the different steps of the algorithm are shown. Also note that there is a circle in picture (d) of the figure to reflect bad quality elements. This is to reflect that morphing an ATLAS into another domain is not an easy task. The ATLAS must be prepared to confront those types of problems. If in most cases, the target meshes differ a lot from one to another in a specific region, the ATLAS can prevent this type of problems by adding points in those zones.

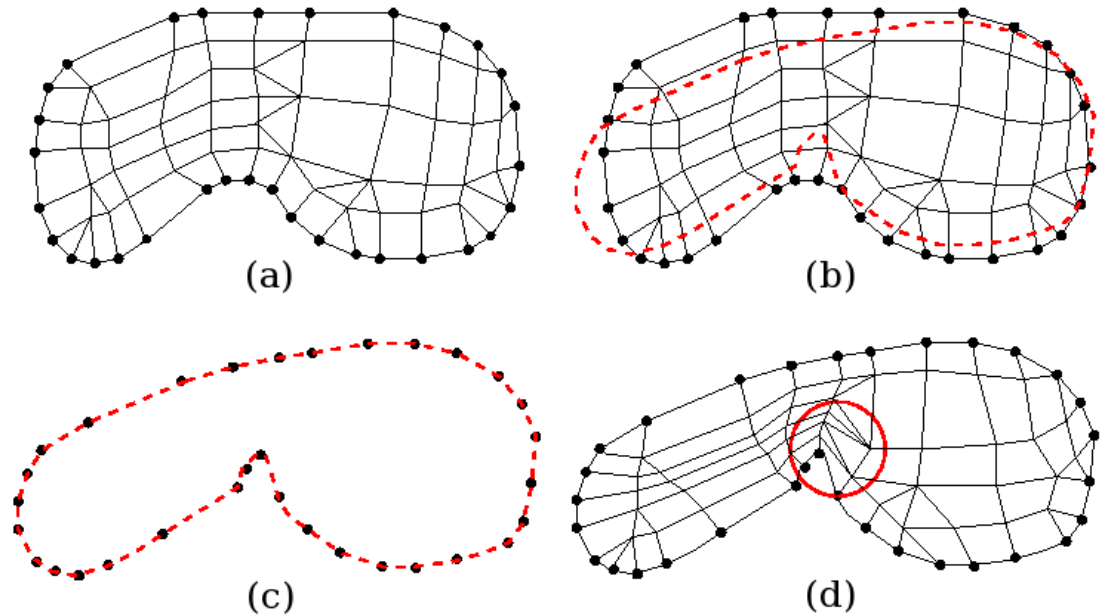

*Figure 18: The mesh-matching algorithm: (a) The ATLAS mesh, (b) the ATLAS and the target surface domain, (c) the matching of ATLAS surface points into the target domain and (d) the final target mesh obtained by applying the transformation function T into the entire ATLAS mesh.*

Another strategy to solve this problem is to use triangles (or tetrahedras in 3D) on those zones. The benefit of this is that contrary to quad (or hexahedra) it is much more difficult to put a triangle or a tetrahedron in an invalid state (concave triangle faces do not exist as in the case of rectangle or hexahedra). Therefore a probably good atlas would be the one shown in figure 19. Note that knowing the zones that deform the most, allows adjusting the number of elements to put in each zone. Therefore, this final mesh represents better the target domain with fewer elements than the resulting one from figure 18d (as fewer elements are presented in the zones where the deformation is less important).

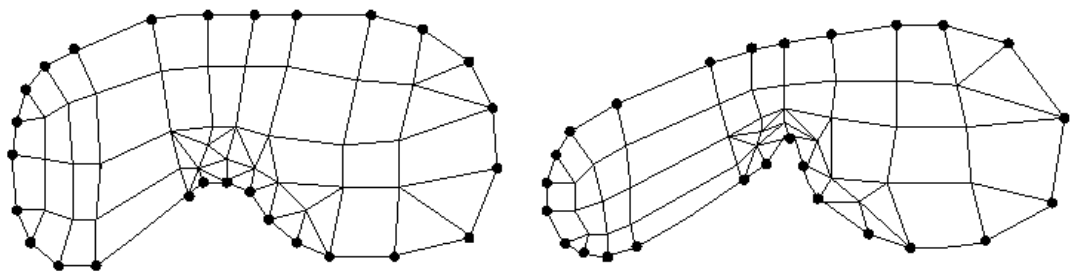

*Figure 19: An improvement to the ATLAS: using more triangles in complicated zones (left) can improve element quality in the resulting mesh (right).*

The M-M algorithm can be used with meshes of any type of elements. In the particular case of Couteau *et al.* (2000), the goal was to obtain femora head meshes and the tested ATLAS was composed of hexahedra and wedges (prisms). In the article, a test over 10 proximal femora is presented. The level of element distortion was satisfactory in comparison to the initial ATLAS.

**Meshing 4D Domains**

The work of Montagnat and Delingette in 2005 is focused on building *surface models* of 3D domains over time (4D). Despite the surface models (instead of volumetric meshes), this work is important to this chapter as it considers deformable domains through time. But before explaining the 4D technique, lets see the 3D deformable meshes from the same authors in 1998.

The ATLAS mesh (initial model), is a simplex mesh. This mesh can be seen as the dual mesh of a triangulation. Each point of the simplex corresponds to a triangle's centroids[6] in the triangulation as shown in figure 20 where the simplex mesh is represented by dashed lines. The simplex mesh has some interesting properties like constant connectivity (each point is connected to three others in the discretization), highly deformable, avoiding surface parametrization problems and other properties as mentioned by Delingette (1999).

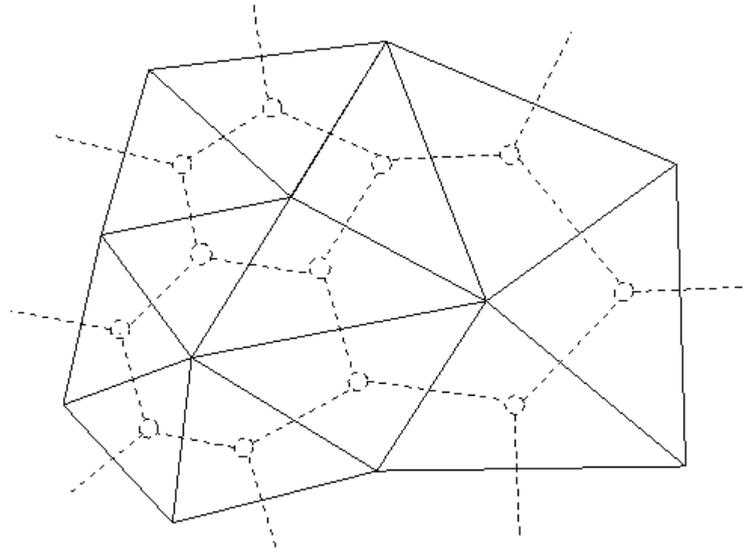

*Figure 20: A triangle mesh and its corresponding simplex mesh (in dashed lines) by the junction of all the triangle's centers.*

Back to the work of Montagnat and Delingette in 1998, an analysis of registration techniques *vs* Free Form Deformations (FFD) is made. It mentions the positive and negative aspects of both techniques and finally reports a workflow to merge them. It analyzes the different results obtained by each technique.

There are several types of registration. The basic one is rigid registration, that considers 6 Degrees Of Freedom (DOF). Then, as the number of DOF increases, the registration can achieve a better representation. The authors work shows results with 6,7 and 12 DOF. It also analyzes the result with one registration governed by cubic B-spline.

---

[6] Note that this is different to the *Voronoi diagram*. If not familiar with this last concept, see: http://en.wikipedia.org/wiki/Voronoi

To couple registration and FFDs, the authors introduce a λ factor that allows the interaction of both registration and FFD. When λ = 0 the point's displacement has a minimum of DOFs (Registration). As λ approaches to the value 1, the number of DOFs increases. Let $V_i^{t+1}$ be the position of vertex *i* in iteration *t+1*, then the position of any vertex in the mesh can be obtained as:

$$V_i^{t+1} = F(V_i^t, V_i^{t-1}) + \lambda f_i^{FFD} + (1-\lambda) f_i^{registration}$$

where F is a function of the previous position of the vertex *i*, $f_i^{registration}$ are the forces that control the regular registration and $f_i^{FFD}$ are the forces that control FFD.

With this equation, it is possible to iteratively increase the DOFs. The first "rigid" registration will stop when a threshold of the level of displacement is achieved. Then the DOFs are incremented by the use of λ and the registration continues until the final registration becomes a FFD with cubic B-splines.

The proposed registration hierarchy allows this work to efficiently match one atlas against a cloud of points as some other optimizations are proposed in the article. But going one step further, some organs in the body are important to analyze in motion. Therefore a substantial application of the presented work is to model 3D domains that change with time (4D), as for example: the heart.

The work of Montagnat *et al*. in 2005, first study a set of images that represent the states of the target organ through time. With the ATLAS of the organ, registration matches the cloud of points of the domain in the different states. As the domain changes, the ATLAS changes too and through all the different states, it simulates the motion of the organ.

## Mesh Warping

As mentioned by Wolberg in 1998, there are several algorithms to morph an image into another one. One of them is the mesh warping algorithm that was first introduced for a special effect sequence in the movie "Willow" in 1988 by Smythe (1990). It allows producing a smooth transition between two images by using the same mesh in both images and identifies some control points that will lead the transformation.

The work of Castellano-Smith *et al*. in 2001 follows the same principle but produces volumetric meshes. It is strongly based on the 2D image registration proposed by Schnabel *et al*. (2001) that works with grids.

Following the same principle as the one in the previous section, the schema of Schnabel *et al*. (2001) is divided in two steps. First, the global motion is corrected using a rigid (6 DOF) or affine (12 DOF) transformation. The global motion then becomes the starting estimate for the second stage, where the local motion is further modeled using Free Form Deformations (FFD) based on B-splines. The combined motion model can be written as:

$$T(x,y,z) = T_{global}(x,y,z) + T_{local}(x,y,z) \quad (Eq.\ 1)$$

with the local motion at each point given by the 3D tensor product of the familiar 1D cubic B-splines. The optimal transformation T is determined by minimizing a registration cost function:

$$C = -C_{similarity}(I_A, T(I_B)) + \lambda C_{deformation}(T) \quad (Eq.\ 2)$$

The similarity term maximizes the voxel similarity between the ATLAS image ($I_A$), and its counterpart in the target image ($I_B$). The deformation cost term is defined as the 3D equivalent of a thin-plate bending energy in order to maximize the smoothness of the transformation, weighted by a factor λ.

As mentioned in the article, single-resolution FFDs are limited by low mesh resolutions, and may develop folding at high resolutions. Therefore a multi-resolution mesh representation is proposed. Each image is registered several times by using grids of different level of refinements. Note that control points can differ from one grid to another one. After registration, the local deformation of each point in the image volume domain is given by the sum of the local deformations across levels:

$$\mathrm{T}_{local}(x, y, z) = \sum_{h=1}^{H} \mathrm{T}^{h}_{local}(x, y, z) \qquad (Eq.\ 3)$$

where each $\mathrm{T}^h_{local}$ is computed with respect to the B-spline of level h. The use of multi-resolution FFDs avoids the folding problem. Therefore the deformation cost term regularizing the smoothness of the deformation in equation 2 is no longer crucial ($\lambda$ can be set to 0).

In practice, this technique is applied by Castellano-Smith *et al.* in 2001. In that study, a volumetric ATLAS of the brain is built from a set of MRI and the control points are selected. To build a new brain mesh, it is necessary to count with MRIs of the target (equivalents to the ones from the ATLAS) and to identify the position of the control points. The registration process then matches the ATLAS mesh to the target mesh. Note that the registration provided in the paper is made from a set of 2D images. As their points are correlated to a 3D mesh, it is possible to obtain a 3D mesh as a result.

The big difference between this method and the Mesh Matching (M-M) technique is that it allows identifying inner points to lead the registration. This is very useful to consider the matching of inner structures like in the case of the brain. However, Berar, Desvignes, Bailly and Payan proposed in 2004 a combined approach using the M-M and a multi-resolution grid for matching high resolution ATLAS into low density target data. This was applied to mandible simulation. Another important remark over this last work is that for the definition of control points, they didn't use geometrical points but anatomical ones like the teeth, even though the teeth were not presented in the target model.

Even though the ATLAS can be built with different meshing techniques, it is important to mention that in the work of Castellano-Smith *et al.* (2001), the ATLAS was built from a set of segmented images. Then with a based Marching-Cubes algorithm they produced a surface model of the brain and inner structures. Their paper doesn't give details on the volumetric mesh generation; however it mentions that the final mesh is a tetrahedral mesh of high quality. The last important thing to mention is that they have used quadratic tetrahedra, i.e. 10 nodes.

Finally the differences between this registration process and the one proposed by Montagnat *et al.* (2005) are:
- The type of mesh they use (grid *vs* simplex).
- The first approach uses an iterative strategy where the level of the mesh refinement and the registration are coupled in each iteration.
- The entire process is leaded by some specific control points that are added in each iteration.

## Untangling and improving quality after mesh adaptation

The accuracy and efficiency of the solution to numerical systems depend on the quality of the mesh. Moreover, in the case of registration techniques, it is possible to compress/stress an element to the point where it becomes folded (i.e. degenerated and therefore invalid for the FEM) as figure 21 shows. This can occur since the registration techniques do not consider the geometry of the elements in the mesh (the registration is made regarding the points and not the elements).

The workflow of element reparation after registration should be:
- Make the mesh valid (repair tangle and folded elements) and
- Improve element's quality when possible.

In the work of Schnabel *et al*. (2001) presented in the previous subsection, this was done in 2D for each image as the idea of hierarchically increase the refinement level of the mesh[7] avoids the presence of "folded" squares. When passing to the 3D volumetric mesh, this problem was not relevant as the final mesh was tetrahedra-only type. Note that folded faces cannot be produced with triangle faces.

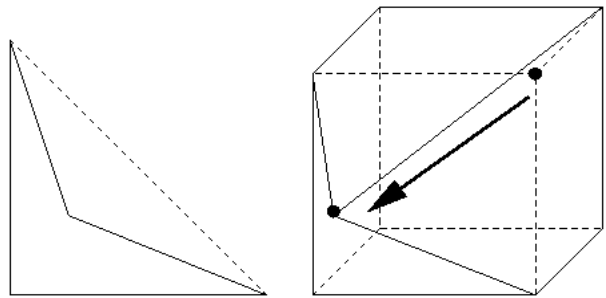

*Figure 21: A folded face(left) where the dotted line is a diagonal and a folded hexahedron (right) where the dotted lines represent a perfect hexahedron.*

One solution to solve tangled and folded elements was proposed by Luboz et al. (2001, 2005). This work continues the Mesh Matching (M-M) technique presented before as it repairs the invalid elements that are sometimes produced after the registration.

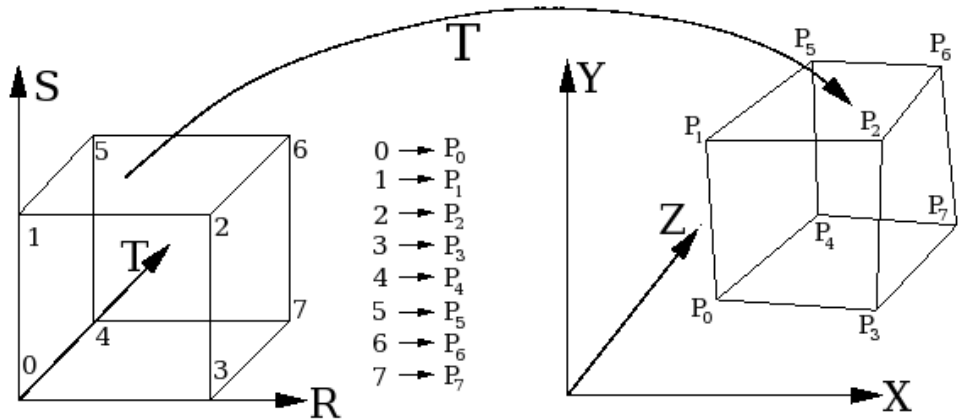

*Figure 22: The reference and actual space from which the Jacobian matrix is calculated. Left: the reference space and right: the actual space.*

The first step consists in detecting the tangle elements. To do so, the **Jacobian matrix** is used regarding the transformation between the reference and the actual space. This transformation can be computed for any type of element. In particular figure 22 shows an example for a hexahedron:

---

[7] Count with several meshes of different levels of detail. For example a grid with 4, 16 and 64 quads.

at the left the reference space $R_{rst}$ and at the right the actual space $R_{xyz}$. The **Jacobian matrix** for this transformation can be written as:

$$J = \begin{bmatrix} J_{xx} & J_{xy} & J_{xz} \\ J_{yx} & J_{yy} & J_{yz} \\ J_{zx} & J_{zy} & J_{zz} \end{bmatrix} = \begin{bmatrix} \sum_i \frac{\partial N_i}{\partial r} x_i & \sum_i \frac{\partial N_i}{\partial r} y_i & \sum_i \frac{\partial N_i}{\partial r} z_i \\ \sum_i \frac{\partial N_i}{\partial s} x_i & \sum_i \frac{\partial N_i}{\partial s} y_i & \sum_i \frac{\partial N_i}{\partial s} z_i \\ \sum_i \frac{\partial N_i}{\partial t} x_i & \sum_i \frac{\partial N_i}{\partial t} y_i & \sum_i \frac{\partial N_i}{\partial t} z_i \end{bmatrix}$$

Where $N_i$ are the interpolation shape functions, $x_i$, $y_i$ and $z_i$ are the actual coordinates and r,s,t are the coordinates in the reference space.

It is possible to compute the Finite Element (FE) if this transformation matrix exists for any element in the mesh. This is equivalent to say that simulation is possible when this matrix is not singular. It is possible to detect singularities by the analysis of the sign of the determinant of *J* *(detJ)*. By evaluating *detJ* in each vertex of the element, a singularity is presented when not all these values have the same sign (i.e. when a null value is presented somewhere inside the element).

In order to repair those elements, the gradient of the *detJ* is computed. For a distorted element *k*, a displacement *Disp*$_{k,j}$ for the vertex *j* is proposed as follows:

$$Disp_{k,j} = \sum_i \nabla_j \text{detJ} = \begin{bmatrix} \sum_i \frac{\partial \text{detJ}_i}{\partial x_j} \\ \sum_i \frac{\partial \text{detJ}_i}{\partial y_j} \\ \sum_i \frac{\partial \text{detJ}_i}{\partial z_j} \end{bmatrix}$$

where *i* is the vertex index of element *k* that has a negative or null value (assuming that all the other vertices of the element have a positive value for *detJ*. The total displacement *Disp*$_j$ for a given vertex *j* is computed by adding each *Disp*$_{k,j}$ obtained for every invalid element *k* where the vertex *j* is presented. This displacement vector is then normalized and the final position *P'*$_j$ for a given point *j* with current position is *P*$_j$ can be obtained as follows:

$$P'_j = P_j + Disp_j * W_j$$

where $W_j$ is a weight factor chosen to constrain the displacement.

Another approach to solve this problem was proposed by Li and Freitag in (1999). This work intends an optimization approach to untangle (equivalent to solve the folding in 2D) the mesh. It searches to maximize the minimal area or volume of the elements. The first step is to detect tangled elements. This is done by computing the Jacobian sign at each element's vertex. A very important point must be noticed here: the work mentions that an element with one vertex of negative Jacobian can be a valid element. Note that this refers to tangling, i.e., an element can be not tangled having one negative Jacobian vertex. However, this has nothing to deal with the FEM. A negative Jacobian vertex does not allow to compute the FEM, therefore the statement should be: a mesh can be untangled but count with negative Jacobian vertex, however this mesh will not be valid from a FEM point of view as it is not possible to compute a simulation with.

The laplacian smoothing technique (Freitag 1997) consists in positioning each vertex *v* of the mesh in the center of the polygon formed by all the vertex connected to it. This method operates heuristically and does not assure quality improvement over the elements of the mesh; for example it can produce slivers[8]. Therefore, optimization techniques based on some quality measure (dihedral angle, aspect-ratio, etc.) are preferred. Unfortunately optimization approaches have a computational cost much higher than the laplacian smoothing. Therefore a combined strategy is proposed by Freitag (1997). But before going into details it is necessary to introduce two concepts.

The laplacian (a smoothing process) by its own does not consider quality measures. Therefore the first modification is the introduction of a "smart laplacian". This is the typical laplacian smoothing but it will be applied only if the new position for *v* improves the quality of the elements.

Even though it is not presented like this, for understanding purposes let's say that the active value *actVal*$_v$ corresponds to the worst element quality associated to vertex *v*, and *actVal*$_{mesh}$ the worst element quality in the mesh. Now to improve element quality, a user-defined *threshold* value is used and the following iterative algorithms are proposed (each algorithm is independent):

1. If *threshold* < *actVal*$_{mesh}$ the *smart laplacian* is performed. Otherwise the laplacian is used.
2. Apply *smart laplacian* and compute the new *actVal*$_v$. If *threshold* < *actVal*$_v$ repair it using laplacian.
3. If *threshold* < *actVal*$_{mesh}$ the process is finished. Otherwise apply *smart laplacian* and compute the new *actVal*$_{mesh}$. If *threshold* < *actVal*$_{mesh}$ apply laplacian.
4. Compute *threshold* = *actVal*$_{mesh}$ + C, where C is an arbitrary constant value. Then proceed as the second algorithm (from this list).

Freitag (1997) concludes that all the above algorithms but the third were superior to the *smart laplacian* and optimization approach. However, the third combined approach was the fastest and it also improved the mesh (only with a worst quality than the others but still, better than the input).

## MESH GENERATION

### Red Green tetrahedral meshing technique

Molino, Bridson, Teran and Fedkiw (2003) have introduced the red green subdivision algorithm. This meshing technique is said to be for modelling highly deformable objects.

The implemented approach starts with a regular octree mesh, i.e. all the octants in the mesh are refined at the same level. Therefore it can be seen as a voxel-based approach but conserving the tree structure of an octree.

Once the desired level of refinement is achieved, each cube is divided in 6 pyramids and each of them in 2 tetrahedrons. As this is done for each hexahedron in the mesh, the global tree structure can be conserved. The difference is that for each leaf in the tree, there will be a reference to the 12 tetrahedrons that replace the original cube.

---

[8] Tetrahedra where all the points tend to be co-planar.

In order to achieve good surface representation some tetrahedrons must be split. Therefore a pattern that enables tetrahedron subdivision is applied over the regions where more elements are needed. This pattern can be applied recursively over tetrahedra. Those elements are said to be *red* (they can still be refined). However, between refined and not refined regions there must be a consistent transition. In order to avoid propagation of the refinement level, new *green* patterns (see figure 23) are applied. Those green tetrahedrons cannot be refined anymore. If for some reason a green tetrahedron must be refined later, it is replaced including his *brothers* with the hierarchical red father (using the tree data structure that enables this process to be fast and robust). Now, with this red tetrahedron the subdivision is applied until the desired element density is achieved.

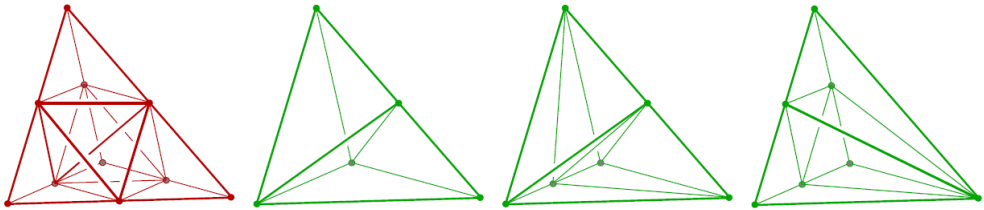

*Figure 23: The red green patterns to split tetrahedrons. The standard red refinement (left) and the three green patterns for transitions between regions of different refinement levels (right).*

The red green meshes are desirable because, except for the green tetrahedrons in the transitions, they produce equilateral tetrahedrons (in difference with a pure *Delaunay* approach), avoiding sliver tetrahedra (which increases the level of computation errors in the simulation). It is for this reason that they are said to be optimized for objects that will be highly deformed.

In the context of neurosurgery and brain modeling, Fedorov, Chrisochoides, Kikinis and Warfield in (2005) proposed to use the FEM to project the output mesh of the *red green* algorithm in order to achieve the correct external surface representation. They defined the displacements of the boundary nodes toward the object surface using a distance map, and solved the system with boundary conditions defined by the displacements of the mesh vertices.

Even though the general context of this work is classified under the real-time simulations, the mesh used to carry out the study should be classified as a mesh for accurate simulations. Note that if this work doesn't use parallel computing, the simulation takes one hour or more to be computed.

## Marching Cubes with Delaunay-based meshes towards several bodies simulation

The work of Audette, Delingette, Fuchs, Burgert and Chinzei (2007) might be considered as an accurate simulation because it builds detailed meshes for several structures of the brain. However their intention is to produce a realistic simulation of the pitatoria surgery for training purposes. Through virtual reality, this simulation is produced in real-time, therefore it can be considered as part of the real-time simulations.

To simulate the endoscopic pituitary surgery, they produce several non connected triangle surface meshes. Each mesh represents a volumetric or a surface mesh of the head structures like the pituitary gland and the arteries, the cranial nerves, the dura mater, the brain and other pathological structures.

In order to produce those surface meshes they first generate a dense surface mesh of high accuracy and topological fidelity, resulting from the Marching Cubes algorithm. The second step corresponds to a decimation process in order to obtain a mesh of the desired density.

The third step consists in building the simplex mesh of the previous calculated mesh. An optimization process leaded by a balloon force can act over the points of the simplex mesh to cause it to expand until some image-based force halts this expansion. In consequence each dual triangle is reallocated in order to count with edges of equal or locally consistent length. In other words, as the relation between a triangle mesh and the simplex is one-to-one, a displacement over a point in the simplex mesh causes a displacement over the three corresponding points in the triangle. Therefore the optimization over all the simplex points improves the quality of the triangles (as an optimization runs over each triangle's center).

Once all the high quality meshes are obtained, Audette *et al.* (2007) propose to add inner points inside the surfaces that will be modeled as a volumetric object. To do so they use a criterion of "near-equal length of the tetrahedron". Once they have enough inner points the authors proceed to build a Delaunay mesh of it.

Finally, they improve the quality of the mesh by making an optimization over the dihedral angle[9]. They achieve an excellent representation of the inner structures of the brain, allowing via Virtual Reality, the training on pitatoria surgery. This strategy is under the category of Volume generation techniques because even though the authors include the input surface mesh to produce the volume mesh, in a previous step, they regenerate the surface mesh with the Marching Cubes technique. Then they build a new dense surface mesh that soon after is decimated and goes through a quality improvement process.

A previous work by Pages, Sermesant and Frey in 2005, developed a general purpose technique similar to the one above. The goal was to develop an application to go from MRI and CT images to volume meshes. The first step was to achieve image segmentation. As this process is even more complicated when the simulation must consider several sub-structures of the image, they used a user-guided segmentation algorithm. Once the segmented images are produced, the only user interaction is to decide the level of precision the mesh must have. With the segmented images, the marching cubes algorithm is applied to produce an initial surface mesh. Then a bi-laplacian algorithm produced a smooth surface (this step wasn't applied by Audette *et al.* in 2007) to finally proceed with a decimation algorithm to decrease the number of triangles. Now with a "good" quality surface mesh, the inner points are inserted. These new points are inserted regarding the Delaunay property to produce an initial volume mesh. In this work, the quality of a tetrahedron, is defined by $Q = \alpha\rho / h$, where $\rho$ is the inradius, $h$ the largest edge length and $\alpha$ is a normalization coefficient (to get a quality of 1 for a regular element). Following this quality function, an optimization process improves the element's quality by node relocation and face flipping operations.

## Variational tetrahedral meshing

This work was developed by Alliez, Cohen-Steiner, Yvinec and Desbrun in 2005. It is a general purpose meshing technique starting from a Delaunay mesh. Then the meshes are improved in quality by optimal point insertion in terms of tetrahedral quality.

Concerning the mesh generation, the important constraints to this work are:

---

[9] The angle formed between two planes.

- Count with a fair shape quality measure. They make a reference to the work of Shewchuk (2002a) and conclude that: "The *radius ratio*, which takes the quotient of inscribed and circumscribed sphere radii (times three for normalization purposes), is a good measure for any kind of degeneracy". This specially avoids the presence of sliver elements.
- Sizing requirement. Citing directly from Alliez *et al.* (2005): "Accuracy and efficiency of numerical solvers depend on the local size of tetrahedra. Consequently, a sizing field, prescribing the ideal local edge length as a function of space, must be added". In consequence, a smooth transition between elements of different sizes is necessary to avoid bad dihedral angles.
- Boundary requirements. Two options are possible to represent the external surface of the domain to mesh: directly include the input surface mesh in the final output mesh or re-mesh the input domain in order to control its quality and density. The first, of course, achieves perfect surface representation as the Advancing Front algorithm, however as Alliez *et al.* (2005) intend to produce a meshing technique for general purposes, the second option was chosen. In this manner, there is no dependency on the density or quality of the input domain.

Regarding all the above points, Alliez and colleagues (2005) present a Delaunay-based optimization technique, called "Variational Tetrahedral Meshing", to efficiently mesh a bounded 3D domain of arbitrary topology or number of connected components. The base of the algorithm considers optimal surface approximation by the work of Cohen-Steiner, Alliez and Desbrun (2004) and Optimal Delaunay Triangulation by the work of Chen and Xu in 2004. Regarding those techniques the authors propose a simple minimization procedure that alternates global 3D Delaunay triangulation and local vertex relocation to consistently and efficiently minimize a global energy over the domain. It results in a robust meshing technique that generates high quality isotropic meshes in terms of radius ratios, as well as angles. A notable feature of the method is that it removes slivers (the principal problem of the Delaunay technique already explained) inside the domain. To provide a flexible meshing tool, they also introduce an automatic sizing field construction that guarantees an arbitrary smooth gradation of the mesh together with faithful approximation of the domain boundary.

The authors provided several comparisons with other quality tetrahedral meshing techniques. In all the presented cases their technique gave the highest mesh quality in terms of tetrahedra radius ratios (the one that avoids sliver tetrahedrons and also considers the dihedral angle).

### A brief remark concerning surfaces

As reviewed in the previous works, to get a good representation of the input domain surface is always a difficult task. Many approaches go through high refinement levels to achieve such a good representation and then, by decimation or point reallocation, decrease the number of faces and try to improve their quality.

Another option to improve surface quality is to change the topology of the mesh. This is not exactly decimation as the work of Bunin in 2006 shows. Even though it is focused in 2D, it can be adapted to improve the quality of surface meshes in 3D. The quality improvement over the surface is achieved by removing "bad quality" or "undesirable" sub-domains of the mesh and by re-meshing those zones. Contrary to decimation, this doesn't only change the topology but can also improve the orientation of the elements. In particular this approach improves quadrilateral meshes.

When the focus is triangle surfaces, one remarkable work is the one by Coll, Guerrieri and Sellares in 2006. The algorithm combines different meshing improvements techniques that incrementally modify an initial triangle mesh by local changes. The element insertion or removal is done in such a way that the mesh quality is maintained during the process. The local operations are: deletion, insertion and addition of Steiner[10] points. The overall process goes through a definition of quality zones. Therefore vertices of skinny triangles are forced to those zones. If a vertex must be removed, new vertices are placed in quality zones to improve the overall quality.

**Hexahedral meshes**

Unstructured quadrilateral/hexahedral meshes are also an important alternative to simulate applications with the Finite Element Method (FEM). However, it still remains a challenging and open problem to generate adaptive and quality quad/hex meshes directly from volumetric data, such as Computed Tomography (CT) and Magnetic Resonance Imaging (MRI).

The work of Zhang and Bajaj in 2006 goes in this direction. The overall strategy consists of:
- Select a starting level of octree refinement for uniform mesh generation with correct topology, regarding the volumetric data.
- Produce a representative quad/hex mesh of the input domain.
- Improve the quality of the final elements.

Once an initial octree level of refinement is achieved five patterns enable transitions between refined and coarse regions. Those patterns are applied in function of the vertex signs. This is based on the "dual contouring method" proposed by Ju, Losasso, Schaefer and Warren in 2002. This last method consists in an approximation of the surface by assigning directions to the vertices in all the sections where domain intersection is produced.

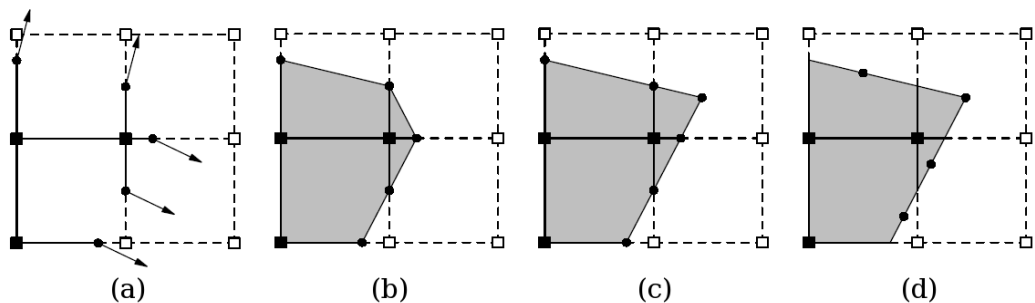

(a)　　　　　(b)　　　　　(c)　　　　　(d)

*Figure 24: Extended Marching Cubes and dual contouring: (a) Octree mesh with domain cutting intersections and the normal of those cutting points, (b) the marching cubes output, (c) the extended marching cubes contour and (d) the dual contour mesh.*

Figure 24 shows the contour detection and assigns a sign to each octree vertex regarding the normals of the intersection points. This can be seen in (a) as the black points rest inside the domain. Image (b) represents the output from the marching cubes technique. Image (c) shows the "extended marching cube" (EMC) technique proposed by Kobbelt, Botsch, Schwanecke and Seidel in 2001 and (d) the points of the dual contour regarding the EMC. As the already introduced triangle simplex mesh, here a dual mesh is expanded to a more global concept as it is not restricted to triangles (see Ju *et al*. (2002) for more details).

---

[10] If in a given Delaunay tetrahedralization or triangulation of an arbitrary domain, a new point is inserted in such a way that the entire mesh still satisfies the Delaunay property, this new point is said to be a *Steiner* point.

With all these techniques, the work by Zhang *et al*. (2006) allows to produce quad/hexa meshes that have regions with different levels of refinement and achieve a "good" surface representation. Finally Zhang and Zhao (2007) continues the previous work by adding two new patterns to drastically decrease the number of transition elements between refined and coarse regions.

## SUMMARY OVER PRESENTED TECHNIQUES

In the two previous sections, several meshing techniques were presented. In the following tables a summary of them is proposed, involving all the relevant aspects of meshing.

**Table 1**

| Meshing Technique | Delaunay | Octree-alike | Marching Cubes |
|---|---|---|---|
| Ruppert (1995) | Ok | - | - |
| Schewchuk (2002a) | Ok | - | - |
| Miller (2003) | Ok | - | - |
| Abell (1999) | - | Ok | - |
| Ashburner (2000) | - | Ok | - |
| Finkel (1974) | - | Ok | - |
| Nesme (2008) | - | Ok | - |
| Lorensen (1987) | - | Ok | - |
| Chernyaev (1995) | - | Ok | Ok |
| Payne (1990) | Ok | Ok | - |
| Ferrant (1999) | Ok | Ok | - |
| Velasco (2007) | Ok | Ok | Ok |
| Couteau (2000) | - | - | - |
| Castellano-Smith (2001) | - | Ok | Ok |
| Berar (2004) | - | - | - |
| Molino (2003) | Ok | Ok | - |
| Audette (2007) | Ok | Ok | Ok |
| Pages (2005) | Ok | Ok | Ok |
| Alliez (2005) | Ok | - | - |
| Chen (2004) | Ok | - | - |
| Zhang (2006 | - | Ok | Ok |

Table 1 presents the reviewed techniques regarding the basic and most common meshing techniques and concepts. Note how the Delaunay property is used by all the techniques that produce tetrahedral meshes (see table 2). It is important to mention that not all the techniques produce volumetric meshes. Some of them are used only to achieve a good surface representation. Another important remark is that several techniques producing tetrahedra, also use the octree-alike methods to produce a primary tessellation.

Table 2 focuses on the type of element produced by the techniques that generate volumetric meshes. Note how mixed elements meshes are by far the less used.

**Table 2**

| Meshing technique | Tetrahedra | Hexahedra | Mixed-elements |
|---|---|---|---|
| Abell (1999) | - | Ok | - |
| Ashburner (2000) | - | Ok | - |
| Nesme (2008) | - | Ok | - |
| Lorensen (1987) | Ok | - | - |
| Payne (1990) | Ok | - | - |
| Ferrant (1999) | Ok | - | - |
| Velasco (2007) | Ok | - | - |
| Couteau (2000) | - | - | Ok |
| Castellano-Smith (2001) | Ok | - | - |
| Berar (2004) | - | - | Ok |
| Molino (2003) | Ok | - | - |
| Audette (2007) | Ok | - | - |
| Pages (2005) | Ok | - | - |
| Alliez (2005) | Ok | - | - |
| Zhang (2006) | - | Ok | - |

There are more generation techniques than registration methods as table 3 shows. Note that this doesn't mean that in medical applications, the generation methods are more used than registration techniques. As generation methods tend to solve general problems there is an important global effort to improve them. In the other hand, registration methods are mostly used in the medical field as the difference from one target to another one makes possible to re-use already generated meshes (ATLAS). Registration methods are also largely used in the medical field due to the complexity of the anatomical structures. The goal isn't just to achieve surface representation but also to conserve some element orientation (alienation) regarding sub-structures (like different types of materials, behavior, physical properties, etc).

**Table 3**

| Meshing type | Reviewed works |
|---|---|
| Registration | Couteau (2000); Montagnat (2005); Castellano-Smith (2001); Berar (2004) |
| Generation | Alliez (2005); Audette (2007); Molino (2003); Abell (1999); Ashburner (2000); Velasco (2007); Pages (2005); Nesme (2008); Lorensen (1987); Payne (1990); Ferrant (1999); Montagnat (2005); Zhang (2006) |

Also note that several tetrahedral generators improve the quality of the input surface mesh and then continue with some point insertion strategies (using the modified input surface mesh) in order to produce the final volume mesh. In other words, techniques that use the input surface mesh, as the Advancing Front, seem to be "not so used" in the medical field.

## CONCLUSION

The goal of this chapter was to show the different possible techniques to produce a suitable mesh for the FEM. Therefore, several related works were omitted as the meshing techniques they used were already explained.

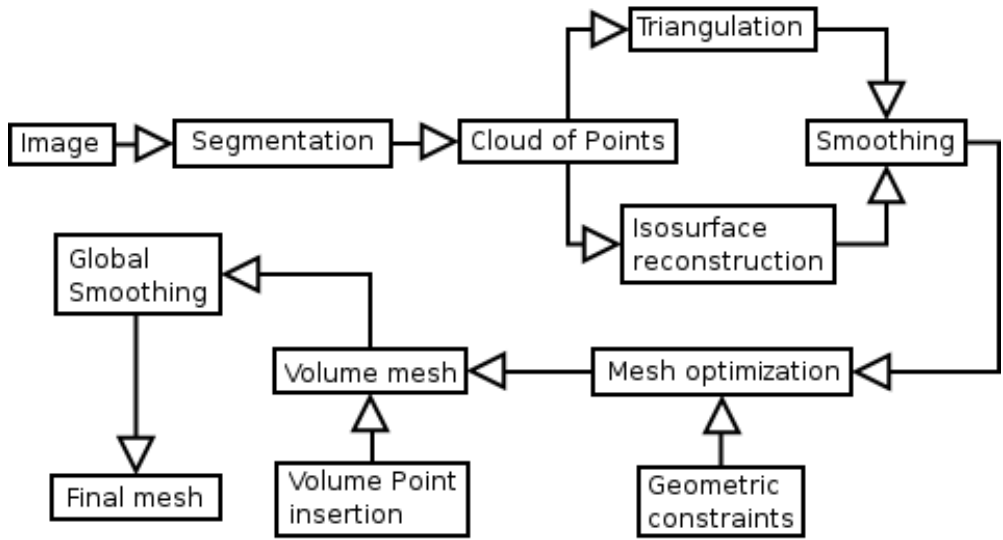

*Figure 25: The mesh generation general flow. From Image to volume mesh generation based on Frey (2004).*

Regarding all the mesh generation techniques presented in this chapter, a workflow for general mesh generation is presented in figure 25. This is based on the work of Frey (2004). In this picture an example of "isosurface reconstruction" is the Marching Cubes algorithm. The "mesh optimization" with the "geometric constraints" refers to the element deletion or insertion per region in function of the simulation requirements.

Finally, regarding all the adaptation techniques, figure 26 shows the equivalent workflow that we propose for a general mesh registration technique. The "rigid registration" can be of different DOF (6,7 or 12) as mentioned above. The "elastic registration" is equivalent to "Free Form Deformation" or "Non Rigid Registration" concepts. The "Element Reparation" corresponds to a re-allocation of the negative Jacobian points and the "quality improvement" needs a measure of quality and then performs an optimization over the elements regarding the selected criterion. Note that a "Smoothing" process is also proposed in order to produce the final volume mesh of the target domain.

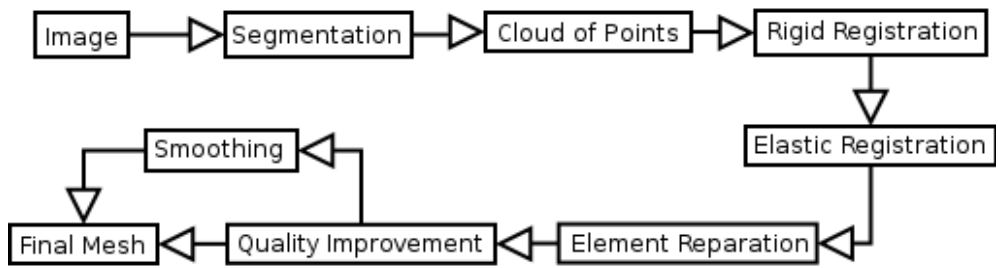

*Figure 26: The mesh adaptation general flow. From Image to volume mesh generation..*

**KEY TERMS & DEFINITIONS**

Continuum Mechanics, Finite Element Method, Mesh, Mesh Adaptation, Mesh Generation, Delaunay property, Grid, Octree, Marching Cubes, Advancing Front and Jacobian Matrix.

**ACKNOWLEDGEMENTS**

Financial support by FONDECYT 1061227, ECOS-Sud C06E04 and Plomo Project (TIC Amsud)